\documentclass[prl,showpacs,preprintnumbers,twocolumn,amsmath,amssymb]{revtex4}
\usepackage{graphicx}
\usepackage{dcolumn}
\usepackage{bm}

\newcommand{\PrOsSb}{PrOs$_4$Sb$_{12}$}
\newcommand{\UThBe}{U$_{1-x}$Th$_{x}$Be$_{13}$}
\newcommand{\UPt}{UPt$_{3}$}

\begin{document}

\title{Pronounced enhancement of the lower critical field and
critical current deep in the superconducting state of \PrOsSb}

\author{T. Cichorek}

\author{A. C. Mota}

\author{F. Steglich}

\affiliation{%
Max Planck Institute for Chemical Physics of Solids, Dresden,
Germany
}%

\author{N. A. Frederick}

\author{W. M. Yuhasz}

\author{M. B. Maple}

\affiliation{%
Department of Physics and Institute for Pure and Applied Physical
Sciences, University of California at San Diego, La Jolla, CA
92093
}%

\date{\today}

\begin{abstract}

We have observed an unexpected enhancement of the lower critical
field $H_{c1}(T)$ and the critical current $I_{c}(T)$ deep in the
superconducting state below $T \approx 0.6$ K ($T/T_{c} \approx
0.3$) in the filled skutterudite heavy fermion superconductor
\PrOsSb. From a comparison of the behavior of $H_{c1}(T)$ with
that of the heavy fermion superconductors \UThBe{} and \UPt, we
speculate that the enhancement of $H_{c1}(T)$ and $I_{c}(T)$ in
\PrOsSb{} reflects a transition into another superconducting phase
that occurs below $T/T_{c} \approx 0.3$. An examination of the
literature reveals unexplained anomalies in other physical
properties of \PrOsSb{} near $T/T_{c} \approx 0.3$ that correlate
with the features we have observed in $H_{c1}(T)$ and $I_{c}(T)$.

\end{abstract}



\maketitle

The filled skutterudite compound \PrOsSb{} has attracted an
enormous amount of interest since it was discovered several years
ago \cite{Bauer02,Maple02}.  This compound is the first heavy
fermion superconductor based on Pr (all of the others are based on
Ce and U), the superconductivity appears to be unconventional, and
the pairing of superconducting electrons may be mediated by
electric quadrupole fluctuations, rather than magnetic dipole
fluctuations that are believed to be responsible for pairing in
the other heavy fermion superconductors.  A number of experiments
have provided evidence for unconventional superconductivity in
\PrOsSb. Structure in the jumps of both the specific heat
\cite{Maple02,Vollmer03,Aoki02} and thermal expansion
\cite{Oeschler03,Oeschler04} associated with the superconducting
transition suggests that there may be two distinct superconducting
phases with superconducting critical temperatures $T_{c1} \sim
1.85$ K and $T_{c2} \sim 1.74$ K in zero field. Superconducting
penetration depth measurements, extracted from muon spin
relaxation ($\mu$SR) experiments in a magnetic field of $200$ Oe
\cite{MacLaughlin02}, and nuclear quadrupole resonance (NQR)
measurements \cite{Kotegawa03} indicate that the superconductivity
of \PrOsSb{} is in the strong coupling regime and has an isotropic
energy gap. In contrast, measurements of the angular ($\phi$)
dependence of the thermal conductivity $\kappa(\phi,H)$ in a
magnetic field $H$ \cite{Izawa03} have been interpreted as
evidence for two distinct superconducting phases, a low field
phase with two point nodes and a high field phase with four or six
point nodes. The superconducting penetration depth $\lambda$,
measured in very low field by means of a microwave technique
\cite{Chia03}, is consistent with point nodes in the energy gap.
Muon spin relaxation measurements in zero field reveal that
spontaneous magnetic moments develop below $T_{c}$, indicative of
time reversal symmetry breaking \cite{Aoki03}. A high field
ordered phase (HFOP), between $4.5$ T and $16$ T and below $1$ K,
has been inferred from electrical resistivity
\cite{Maple02,Ho02,Ho03}, specific heat \cite{Vollmer03,Aoki02},
thermal expansion \cite{Oeschler03,Oeschler04}, magnetization
\cite{Ho03,Tenya03,Tayama03}, and magnetostriction
\cite{Oeschler04} measurements. From neutron diffraction
measurements at high magnetic fields, the HFOP was identified with
antiferroquadrupolar order \cite{Kohgi03}.  This suggests that the
unconventional superconductivity in \PrOsSb{} may occur in the
vicinity of a quadrupolar quantum critical point (QCP), similar to
the situation with certain Ce and U compounds where
superconductivity is found in the vicinity of an antiferromagnetic
(AFM) QCP \cite{Mathur98}.  In this Letter, we report measurements
of the lower critical field $H_{c1}(T)$, critical current
$I_{c}(T)$, ac magnetic susceptibility $\chi_\mathrm{ac}(T)$, and
specific heat $C(T)$ in order to obtain more information about the
unconventional superconductivity exhibited by this intriguing
material.  Our measurements indicate that a transition to another
superconducting phase, characterized by enhanced $H_{c1}(T)$ and
$I_{c}(T)$, occurs deep within the superconducting state at $T
\approx 0.6$ K ($T/T_{c} \approx 0.3$) in zero field.

The \PrOsSb{} single crystals studied in this investigation were
grown from an Sb flux in a manner described elsewhere
\cite{Bauer01}. Powder x-ray diffraction studies of crystals grown
in this run revealed that the samples are single phase.  The
residual resistivity (at a temperature right above $T_{c}$) of
crystals grown in this manner is typically less than
$5~\mu\Omega$cm.  Specific heat measurements were made in a
semi-adiabatic $^3$He calorimeter by means of a standard heat
pulse technique.  The lower critical field $H_{c1}$ was determined
from isothermal magnetization curves taken with a custom made
SQUID magnetometer. In this arrangement, the detection loop is
located in the mixing chamber of a dilution refrigerator, and the
sample is stationary and in direct contact with the liquid
$^3$He--$^4$He mixture.  The ac magnetic susceptibility was
measured in the same arrangement using a mutual inductance bridge
with the SQUID as a null detector \cite{Dumont02}.

\begin{figure}[tbp]
\begin{center}
\includegraphics[angle=270,width=3.5in]{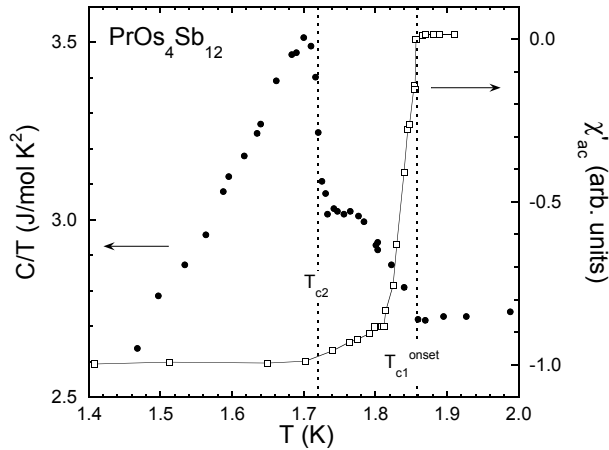}
\end{center}
\caption{Specific heat $C$ divided by temperature $T$ (closed
circles, left scale) and ac magnetic susceptibility
$\chi_\mathrm{ac}$ (open squares, right scale) vs $T$ for the
single crystal of \PrOsSb{} studied in this work.  The value
$\chi_\mathrm{ac} = -1$ has been taken based on the values of
$\chi_\mathrm{ac}$ for $T \rightarrow 0$.} \label{C+chi}
\end{figure}

Measurements of $C(T)$ and $\chi_\mathrm{ac}(T)$ were performed on
the same \PrOsSb{} single crystal in the vicinity of the
superconducting transition.  The $C(T)$ data are shown in Fig.\
\ref{C+chi} and reveal the ``double jump'' structure, reminiscent
of two distinct superconducting transitions at the critical
temperatures $T_{c1} = 1.86$ K (onset) and $T_{c2} = 1.72$ K.  It
is interesting to note that the jump $\Delta{}C_{1}$ at $T_{c1}$
is rather broad, while the jump $\Delta{}C_{2}$ at $T_{c2}$ is
very sharp. Also shown in Fig.\ \ref{C+chi} are
$\chi_\mathrm{ac}(T)$ data, taken with a field amplitude
$H_\mathrm{ac} = 1.2$ mOe and a frequency $f = 160$ Hz. From the
in-phase component of the susceptibility, $90\%$ of the transition
occurs at $T_{c1}$ with the last $10\%$ ``foot'' extending to
$T_{c2}$. At $T_{c2}$, the magnitude of $\chi_\mathrm{ac}$ reaches
its maximum value observed in the limit $T \rightarrow 0$ K.

The ``double jump'' feature in $C(T)$, originally reported in
references \cite{Maple02,Vollmer03,Oeschler03}, has been confirmed
by several other groups \cite{Aoki02,Rotundu04,Measson04} and
seems to be an intrinsic property of \PrOsSb{}. Measurements in a
magnetic field indicate that $T_{c1}$ and $T_{c2}$ track each
other and lie on curves with similar shapes displaced from one
another in the $H-T$ plane \cite{Vollmer03,Aoki02,Measson04}.
Since the large diamagnetic change in $\chi_\mathrm{ac}$ occurs at
$T_{c1}$ due to induced supercurrents, it is clear that the
transition at $T_{c1}$ is associated with superconductivity. The
sharp jump in $C(T)$ at $T_{c2}$ suggests that the transition at
$T_{c2}$ is also due to superconductivity.

\begin{figure}[hbp]
\begin{center}
\includegraphics[angle=270,width=3.5in]{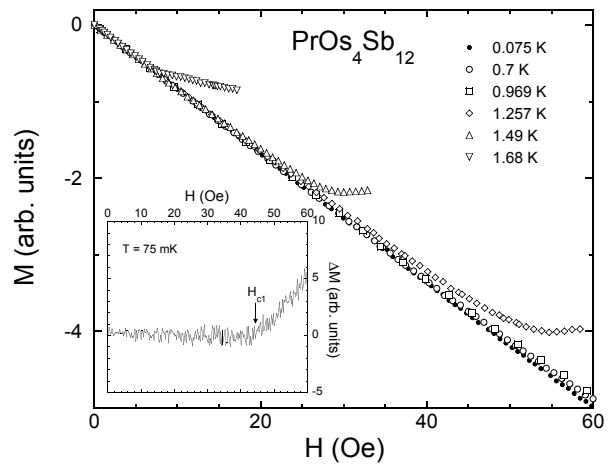}
\end{center}
\caption{Isothermal magnetization $M(H)$ curves taken after
zero-field cooling at different temperatures for the same single
crystal of \PrOsSb{} upon which the $C(T)$ and
$\chi_\mathrm{ac}(T)$ data shown in Fig.\ \ref{C+chi} were taken.
The inset shows the deviation in the $M(H)$ curve at $H_{c1}$.}
\label{M(H)}
\end{figure}

\begin{figure}[hbp]
\begin{center}
\includegraphics[width=3in]{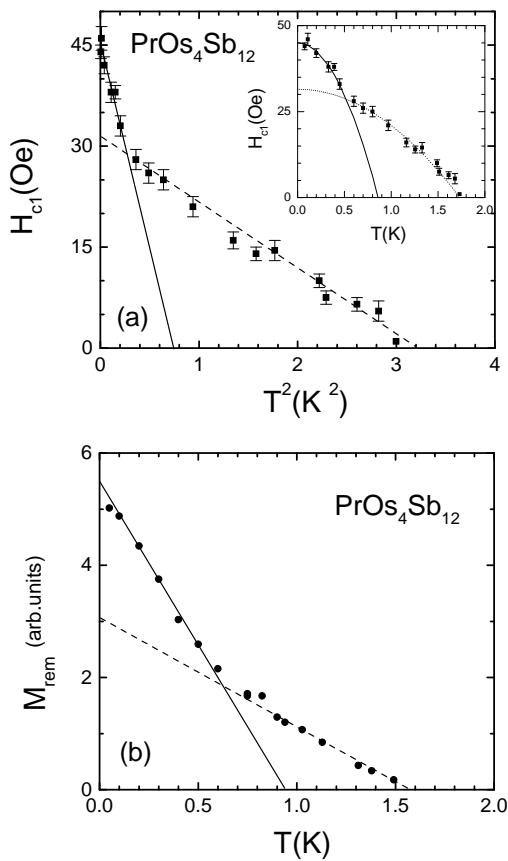}
\end{center}
\caption{(a) Lower critical field $H_{c1}$ vs $T^2$ for the
\PrOsSb{} single crystal. The inset shows the same data plotted as
$H_{c1}$ vs $T$. (b) Remanent magnetization $M_\mathrm{rem}$ vs
$T$ for the \PrOsSb{} single crystal.  $M_\mathrm{rem}(T)$ is
proportional to the critical current $I_{c}(T)$.}
\label{Hc1}
\end{figure}

The lower critical field $H_{c1}(T)$ was determined from
magnetization (shielding) isotherms; typical examples at different
temperatures are shown in Fig.\ \ref{M(H)}.  Each magnetization
curve was taken after zero-field cooling the sample to the desired
temperature. The lower critical field $H_{c1}$ was defined as the
first deviation from the shielding slope in the $M(H)$ curve, as
illustrated in the inset of Fig.\ \ref{M(H)}.  The resultant
$H_{c1}(T)$ data are plotted in Fig.\ \ref{Hc1}(a) vs T (inset)
and $T^2$. We observe a pronounced enhancement of $H_{c1}$ below
$T \approx 0.6$ K. Similar enhancements of $H_{c1}(T)$ have been
observed by various groups in \UThBe{} \cite{Rauchschwalbe87} for
Th concentrations x between $2\%$ and $4\%$ and in \UPt{}
\cite{Amann98} below their second (lower) superconducting
transition temperatures $T_{c2}$. We have fitted the $H_{c1}$ vs
$T^{2}$ data in Fig.\ \ref{Hc1}(a) with two straight lines.
Extrapolation of these lines to $T = 0$ K yields a value of
$H_{c1}(0)$ of $31$ Oe for the high temperature ($T > 0.6$ K)
superconducting phase and $45$ Oe for the low temperature ($T <
0.6$ K) superconducting phase. Since our crystal was in the shape
of a rectangular parallelopiped of $5 \times 0.2 \times 0.3$
mm$^3$ and the magnetic field was aligned parallel to the largest
dimension, we have not introduced demagnetization corrections to
the given values of $H_{c1}$. The sharp kink and enhancement of
$H_{c1}(T)$ reported in this Letter are consistent with the
positive curvature in $H_{c1}(T)$ deduced from previous
magnetization measurements on \PrOsSb{} \cite{Ho03}, although
these measurements did not have enough resolution to reveal the
sharp kink in $H_{c1}(T)$ at $T \approx 0.6$ K.

In Fig.\ \ref{Hc1}(b), we show values of the remanent
magnetization $M_\mathrm{rem}$ obtained by cycling the
zero-field-cooled crystal up to the field corresponding to the
critical state (full penetration of vortices in the sample),
removing the magnetic field, and recording the number of expelled
vortices with a digital flux counter as the crystal is heated to
$T \gg T_{c}$. In this case, $M_\mathrm{rem}$ is proportional to
the critical current $I_{c}$. Coincident with the enhancement of
$H_{c1}$ at $T = 0.6$ K, we observe a dramatic increase in $I_{c}$
below the same temperature, indicating that the superconducting
phase below $T \approx 0.6$ K has substantially stronger pinning.
By comparison with \UPt{} and thoriated UBe$_{13}$, one is tempted
to identify the temperature $T = 0.6$ K below which $H_{c1}(T)$
and $I_{c}(T)$ are enhanced with a third superconducting
transition at a critical temperature $T_{c3} \approx 0.6$ K.

One might ask whether inclusions of free Os in the single crystal
could be responsible  for the enhancement of $H_{c1}(T)$ and
$I_{c}(T)$ below $T_{c3} \approx 0.6$ K, since the $T_{c}$ of pure
Os of $0.66$ K \cite{Hulm57} is close to the value of $T_{c3}$.
However, it seems very unlikely that these features in $H_{c1}(T)$
and $I_{c}(T)$ are due to free Os since x-ray diffraction and
electron microprobe studies do not give any indication of the
presence of free Os in the \PrOsSb{} single crystal within the
limits of detectability.  Furthermore, although $H_{c1}(T)$ of
pure Os is $\sim 1.4$ times larger than that of \PrOsSb{} below
$T_{c3}$ \cite{Hulm57}, it would not be detected in measurements
of $M(H)$ shielding isotherms.  In this case, the first deviation
of $M(H)$ from linear Meissner behavior, corresponding to the
first vortex penetration, would occur at the lowest $H_{c1}$,
namely that of \PrOsSb. Also, inclusions of free Os would not be
expected to increase the pinning of vortices, and, in turn,
$I_{c}(T)$, when they become superconducting.

The second critical temperature $T_{c2}$ inferred from the
structure in $C(T)$ (see Fig.\ \ref{C+chi}) does not seem to be
associated with the critical temperature $T^{*}$ separating two
superconducting phases, a low field phase with two point nodes in
the energy gap and a high field phase with four or six point nodes
in the energy gap, inferred from the $\kappa(\phi,H)$ measurements
in a magnetic field \cite{Izawa03}. The transition between these
two superconducting phases with different order parameter
symmetries has not been observed in any other physical properties
to date.

Zero field muon spin relaxation measurements reveal the
spontaneous appearance of magnetic moments below $T_{c1}$,
indicative of breaking of time reversal symmetry in the
superconducting state. This suggests that the superconducting
phases below $T_{c1}$ and $T_{c2}$ are ``nonunitary'' spin triplet
(odd parity) superconducting states, which have nonvanishing spin
moments \cite{Aoki03}. One possibility is that the two
superconducting phases have different order parameter symmetries,
as occurs in \UPt.  This material has two superconducting jumps in
the specific heat in zero field separated by $\approx 0.05$ K.  On
the other hand, recent microwave surface impedance measurements on
\PrOsSb{} down to $1.2$ K have been interpreted in terms of
Josephson-coupled two-band superconductivity, implying two order
parameters of the same symmetry \cite{Broun03}.  It is clear that
the superconducting phases associated with $T_{c1}$ and $T_{c2}$
are not well understood and that further research is needed to
elucidate their nature.

Other physical properties of \PrOsSb{} have been reported that
exhibit anomalous behavior in the superconducting state in the
vicinity of $0.6$ K ($T/T_{c} \approx 0.3$), where a transition to
a third superconducting phase appears to occur. Sb nuclear
quadrupole resonance measurements \cite{Kotegawa03} reveal that
the inverse nuclear spin lattice relaxation time $1/T_{1}$ does
not exhibit a coherence peak near $T_{c}$, decreases exponentially
with decreasing temperature by over three orders of magnitude, and
then abruptly levels off below $T \approx 0.6$ K. The absence of a
coherence peak near $T_{c}$ is also found in other Ce and U heavy
fermion superconductors. However, the exponential decrease in
$1/T_{1}$ with decreasing $T$ is in marked contrast to the $T^{3}$
variation of $1/T_{1}$ at low temperatures generally observed in
Ce and U heavy fermion superconductors, as expected for line nodes
in the energy gap. This experiment also yields a large value of
the energy gap $2\Delta/k_{B}T_{c} \approx 5.2$ indicative of
strong coupling, consistent with the large value
$\Delta{}C/\gamma{}T_{c} \approx 3$ determined from specific heat
measurements \cite{Vollmer03}. The exponential dependence of
$1/T_{1}$ implies that the superconducting energy gap is
isotropic; interestingly, $\mu$SR measurements of the penetration
depth $\lambda$ in a magnetic field of 200 Oe also yield evidence
for an isotropic energy gap \cite{MacLaughlin02}. In contrast, the
$\kappa(\phi,H)$ data of Izawa et al.\ \cite{Izawa03} and the
$\lambda(T)$ data of Chia et al.\ \cite{Chia03} indicate that
there are point nodes in the energy gap. However, the abrupt
levelling off of $1/T_{1}$ at $T \approx 0.6$ K following its
exponential decrease suggest a transition to a superconducting
phase below $0.6$ K with states in the energy gap.

The measurements of $\lambda(T)$ on a single crystal of \PrOsSb{}
by Chia et al.\ \cite{Chia03} also exhibit a feature in the
vicinity of $T = 0.6$ K. In this study, a small upturn in
$\Delta\lambda$ was observed at $T = 0.62$ K in the three
directions $a$, $b$, and $c$, at which the ac field was applied.
The measured drop in $\Delta\lambda$ at $T = 0.62$ K from the high
temperature values, although rather small, clearly points to an
increase in the superfluid density for $T < 0.6$ K. The
$\lambda(T)$ data were then fitted by the authors from $0.1$ K to
$0.55$ K with power laws of the form $\Delta\lambda(T) = A +
BT^{n}$ with $n \approx 2$, suggesting the presence of low lying
excitations in this temperature range, incompatible with an
isotropic superconducting gap.

A direct measurement of the superconducting energy gap of
\PrOsSb{} was made using a high resolution scanning tunneling
microscope (STM) by Suderow et al.\ \cite{Suderow04}.
Measurements on parts of the sample yielded a superconducting
density of states with a well-defined energy gap and no low energy
excitations.  A plot of the energy gap $\Delta$ vs $T$ has an
overall shape that is consistent with the BCS theory, but with a
small feature near $0.6$ K that could be associated with the
anomalies we have observed at the same temperature in $H_{c1}(T)$
and $I_{c}(T)$. Furthermore, measurements on other parts of the
sample revealed spectra with a finite density of states at the
Fermi level in the superconducting gap. It is possible that the
superconducting phase that appears to form below $T/T_{c} \approx
0.3$ is an inhomogeneous phase consisting of regions with a full
gap and regions with states in the gap.

In summary, we have observed an unexpected enhancement of
$H_{c1}(T)$ and $I_{c}(T)$ below $T/T_{c} \approx 0.3$ in the new
filled skutterudite heavy fermion superconductor \PrOsSb.  From a
comparison of the behavior of $H_{c1}(T)$ with that of the heavy
fermion superconductors \UThBe{} and \UPt, we speculate that the
enhancements of $H_{c1}(T)$ and $I_{c}(T)$ in \PrOsSb{} reflect a
transition into another superconducting phase at $T_{c3} \approx
0.6$ K. An examination of the literature revealed unexplained
anomalies in other physical properties around $T \approx 0.6$ K.
These anomalies should be investigated further in light of the new
observations reported in this Letter.  Surprisingly, there is no
evidence of a jump in the specific heat around $T \approx 0.6$ K
\cite{Maple02,Vollmer03,Aoki02}, even in recent more detailed
measurements \cite{Grube04}.  One reason could be that the
transition at $T_{c3}$ is of first order, like the one between the
A and B phases of superfluid $^3$He, or of a higher order than
second.  A small feature in $C(T)$ could also be obscured by the
nuclear Schottky contribution which increases rapidly with
decreasing temperature at low temperatures below $\sim 0.6$ K.
Finally, the discrepancy between different experiments on single
crystals at $H = 0$, concerning the nature of the superconducting
gap, can be reconciled if the temperature interval covered in the
analysis is taken into account.  Indeed, the NQR anaylsis
\cite{Kotegawa03}, consistent with an isotropic gap was performed
for $T \geq 0.6$ K, while the penetration depth analysis
\cite{Chia03}, consistent with nodes in the gap, was done for $T <
0.55$ K.  In view of the enhancement of $H_{c1}$ and $I_{c}$ below
$T_{c3}$ reported here, it seems plausible that the nature of the
gap function changes at $T_{c3}$.

\section*{Acknowledgements}

Research at MPICPS was partially supported by the Fonds des
Chemischen Industrie, while research at UCSD was supported by the
U.S. National Science Foundation (Grant No. DMR-0335173) and the
U.S. Department of Energy (Grant No. DE-FG02-04ER46105).

\end{document}